\documentclass[conference]{IEEEtran}
\IEEEoverridecommandlockouts
\usepackage{cite}
\usepackage{amsmath,amssymb,amsfonts}
\usepackage{algorithmic}
\usepackage{graphicx}
\usepackage{textcomp}
\usepackage[table,xcdraw]{xcolor}
\usepackage{balance}
\def\BibTeX{{\rm B\kern-.05em{\sc i\kern-.025em b}\kern-.08em
    T\kern-.1667em\lower.7ex\hbox{E}\kern-.125emX}}
\begin{document}

\title{Where’s That Voice Coming?\\ Continual Learning for Sound Source Localization}

\author{\IEEEauthorblockN{Yang Xiao and Rohan Kumar Das}
\IEEEauthorblockA{\textit{Fortemedia Singapore, Singapore} \\
Email: \{xiaoyang, rohankd\}@fortemedia.com}}

\maketitle

\begin{abstract}
Sound source localization (SSL) is essential for many speech-processing applications. Deep learning models have achieved high performance, but often fail when the training and inference environments differ. Adapting SSL models to dynamic acoustic conditions faces a major challenge: catastrophic forgetting. In this work, we propose an exemplar-free continual learning strategy for SSL (CL-SSL) to address such a forgetting phenomenon. CL-SSL applies task-specific sub-networks to adapt across diverse acoustic environments while retaining previously learned knowledge. It also uses a scaling mechanism to limit parameter growth, ensuring consistent performance across incremental tasks. We evaluated CL-SSL on simulated data with varying microphone distances and real-world data with different noise levels. The results demonstrate CL-SSL's ability to maintain high accuracy with minimal parameter increase, offering an efficient solution for SSL applications. 
\end{abstract}


\section{Introduction}
\label{sec:intro}
Sound source localization (SSL)~\cite{ssl1} aims to identify the origins of speech sources by analyzing signals captured through an array of microphones. At the cornerstone of SSL lies the direction of arrival (DOA) estimation, a process crucial for calculating the angles at which sounds reach the microphone array. This spatial information is necessary for speech-related applications~\cite{ss,tse,asr,se} such as speech enhancement and far-field automatic speech recognition. These applications rely on the precise localization of speech sources to significantly improve their performance and functionality. Consequently, the development of more sophisticated SSL techniques has become a key focus within the research community to achieve greater robustness and adaptability in real-world settings. 

Classical signal processing techniques like generalized
cross-correlation (GCC)~\cite{gcc}, multiple signal classification (MUSIC)~\cite{music}, and steered response power (SRP) based methods, particularly SRP-PHAT~\cite{srp}, which is obtained from the SRP by applying phase transform (PHAT) whitening. They are available to adapt to different microphone array configurations. However, their effectiveness decreases in noisy and reverberant environments compared with deep learning solutions. 

A significant milestone in this exploration is the adoption of deep learning architectures such as convolutional neural networks (CNN)~\cite{cnn1,cnn2,cnn3,cnn4,chan22_l3das}. However, deep learning models used for SSL face a unique challenge: the misalignment between training and testing environments, especially concerning the microphone array's configuration and acoustic environments. Such mismatches precipitate a decline in model efficacy, necessitating re-training for new configurations. Re-training always requires extra time and computation resources. This highlights an essential area for ongoing deep learning-based SSL research, to enhance model accuracy across diverse acoustic settings. Although some methods~\cite{bs1,ft3} adapt to the variable array setups relying on specific spatial features, they do not address more complex acoustic configurations like the difference absorption and reverberation time. Along the diverse configurations, one fixed method faces forgetting because the new knowledge overrides the existing knowledge. This problem is known as catastrophic forgetting.

Recently, continual learning (CL)~\cite{cl} has received notable attention for solving catastrophic forgetting. Various works recently incorporated CL to continuously learn new knowledge while retaining previously learned knowledge in speech processing~\cite{cl1,cl2,til}. We can classify existing CL methods into two categories based on whether to use a replay buffer, which stores extra data: exemplar-free methods and replay-based methods. Exemplar-free methods protect parameters learned from previous tasks through loss regularization or dynamic modules without access to any historical data in existing tasks. Replay-based methods store historical data as replay exemplars in a memory buffer. Since SSL applications are typically deployed on edge devices with limited memory, we propose using exemplar-free methods~\cite{pnn,cl3,ucil}. These methods allow SSL to gradually learn new information using simple, widely adopted features like the short-time Fourier transform (STFT) while retaining valuable knowledge from existing acoustic environments and configurations.

In this research, we introduce CL-SSL, a novel {\it exemplar-free continual learning method} for SSL, crafted to solve the issue of adaptation from source to target configurations and environments. The core of CL-SSL's architecture is {\it task-specific sub-networks }, each engineered to memorize knowledge from previous configurations or environments, enabling smooth transitions across diverse acoustic environments while maintaining accuracy on previous tasks. A scaling mechanism is integrated into these sub-networks to limit parameter growth by reducing the extra layers. Contrary to traditional finetuning methods, CL-SSL utilizes a collection of pre-trained sub-networks, establishing lateral connections that extract the crucial features for new tasks, thereby enriching feature presentation. To the best of our knowledge, this is the first work to apply continual learning techniques to adapt SSL to various acoustic configurations. Demonstrated through extensive testing on the simulated and real-world data, CL-SSL exhibits competitive performance on unfamiliar configurations and acoustic environments with less parameter increase and no reliance on previous data buffers. 


\begin{figure*}[t]
\centering  
\includegraphics[width=0.9\linewidth]{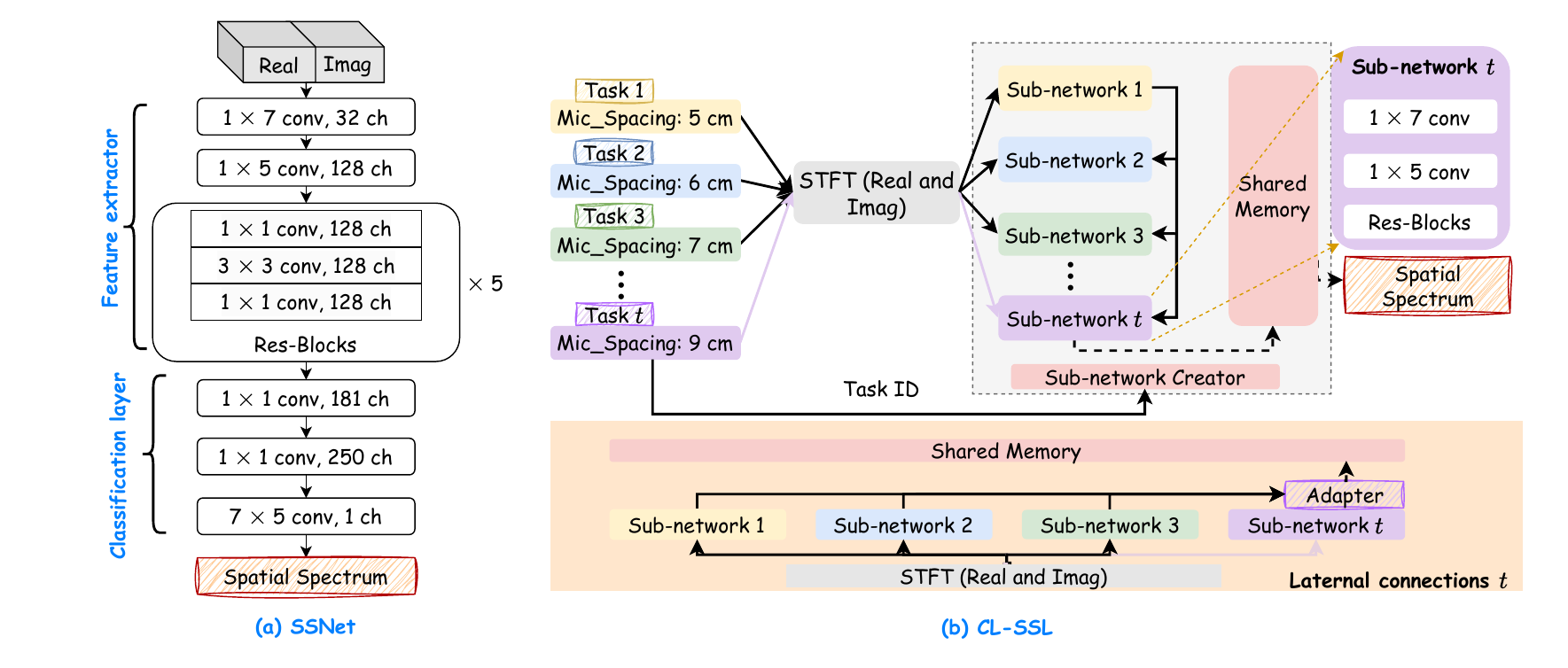}
\vspace{-5mm}
\caption{(a) Architecture of SSNet (b) Proposed CL-SSL framework.}
\label{fg1}
\vspace{-3mm}
\end{figure*}
\section{Related Works}
Recently, several studies have explored the application of continual learning in speech-processing tasks. In~\cite{chencl}, the authors proposed a hyper-gradient-based exemplar strategy for task-oriented dialogue systems, which selects key exemplars for periodic retraining. Xiao et al.~\cite{ucil} developed an independent unsupervised learning framework with a distillation loss to incorporate new sound classes while maintaining the consistency of sound event detection across incremental tasks. The authors of~\cite{peng2024dark} introduced incremental spoken keyword spotting, using model predictions to distill past experiences throughout training. However, these methods do not address the issue of catastrophic forgetting in SSL tasks. In addition, most approaches rely on exemplars from existing tasks, making them less suitable for on-device applications with memory constraints. In a CL-based SSL task reported in~\cite{qian2024analytic}, the authors treat azimuth angles as discrete classes, which may not reflect in real-world scenarios as splitting the range \(360^\circ\) into multiple tasks creates artificial constraints and reduces practical applicability.
\section{Proposed Method}
\label{sec:method}

\subsection{Continual learning for sound source localization}
 In this work, applying continual learning to SSL involves receiving a series of \( T \) distinct tasks (under different configurations and environments) in sequence and optimizing the performance across all tasks with less catastrophic forgetting. For each task \( \tau_t \), with \( t \leq T \), we handle a set of training data consisting of pairs \( (x_t, y_t) \), where \( x_t \) represents the acoustic signals and \( y_t \) are the corresponding direction labels, sampled from a distribution \( D^T \).

Our goal is to minimize the cumulative loss \( L_{tot} \), which aggregates the sub-losses \( L_{ssl} \) for all \( T \) tasks, formulated as:
\begin{equation}
     L_{tot} = \sum_{t=0}^{T} E_{(x_t,y_t) \sim D^T}  [L_{ssl} (F_t (x_t; \mu_t), y_t)]
\end{equation}

where \( L_{ssl} \) represents a classification loss function, such as mean squared error (MSE) loss, and \( F_t(x_t; \mu_t) \) is the SSL model parameterized by \( \mu_t \). During the training, we aim to find the optimal parameters \( \mu_t' \) that perform well across all \( T \) tasks. This is challenging since parameters learned from previous task \( \tau_t \) are prone to be overwritten when learning the subsequent task \( \tau_{t+1} \), leading to catastrophic forgetting.

\subsection{Network backbone for sound source localization}

We adopt a spatial spectrum prediction module using the spatial spectrum network (SSNet) \cite{ssnet1,ssnet2,ssnet3} for SSL. We consider two microphone linear arrays as an instance in this section, although our proposed method can be employed for any number of microphone positions. This system learns complex relationships from audio data for predictions, where the sound could be coming from at every degree in a half-circle of the linear microphone array. To perform this, we combine sound segments from two microphones and transform them into complex STFT. We consider the real as well as the imaginary part of the complex STFT instead of the phase and power and then combine this data to form a feature representation \( v \). This \( v \) feature is then used as the input to SSNet for predicting the spatial spectrum \(\hat{p}\). The network structure of SSNet is shown in Fig. \ref{fg1} (a) which depicts SSNet includes layers to simplify and analyze sound features. It uses two convolutional layers to reduce frequency details and five residual blocks to pick up intricate sound features. Then, it has a layer that focuses on the SSL and rearranges the sound information in preparation for the final step. The result spatial spectrum is an 181-point map.

Following the work in~\cite{ssnet3}, we use a probability-based method to determine the likelihood of a sound coming from any of the 180 directions. This part of SSNet assigns a probability to each direction based on how closely it matches the ground truth direction. Specifically, each element of the encoded \( 181 \)-dimensional vector \( p(\theta_i) \) is assigned to a particular azimuth direction \( \theta_i \in \{0^\circ, 1^\circ, \dots,  180^\circ\} \). The values of the vector follow a Gaussian distribution that maximizes at the ground truth direction, which is defined as follows:
\begin{equation}
p(\theta_i) = 
\begin{cases}
\exp\left( -\frac{d(\theta_i, \theta_y)^2}{2\sigma^2} \right) , & \text{if } |\Theta| = 0 \\
1, & \text{if } |\Theta| = 1 
\end{cases}
\end{equation}
where \( |\Theta| \) is the number of sources, \( \theta' \) is the ground truth direction of one source, \( \sigma \) is a predefined constant that controls the width of the Gaussian function and \( d(\dots,\dots) \) denotes the angular distance. We then train SSNet by computing the MSE loss between the segment-wise output spatial spectrum from the model and the ground truth. In this work, we only consider the single-source localization.

\subsection{Proposed CL-SSL approach}
Inspired by the progressive neural networks~\cite{pnn}, we present a novel exemplar-free continual learning approach tailored for deep SSL, depicted in Fig. \ref{fg1}(b). Our method, termed CL-SSL, is composed of two principal elements: a task-specific sub-network creator for individual tasks and a lateral connection set bridging a suite of sub-networks. To manage the model size, we've introduced a gap-aware layer scaling mechanism in CL-SSL, ensuring the parameter count remains controlled.

\textbf{A task-specific sub-network creator} is designed to generate the \(t^{{\text {th}}}\) sub-network for each new task \(\tau_t\). When a new acoustic configuration SSL \(\tau_t\) comes, the creator adds a sub-network with random initialization which is only the feature extractor part of SSNet. All sub-networks share the common memory which is the classification head of SSNet for a fixed class number 181.

\textbf{A lateral connection set} serve a pivotal role. When the \(t^{{\text {th}}}\) sub-network processes input data, it not only forwards its output to the classification layer but also receives output features from all previously established sub-networks. To integrate these multiple outputs effectively, each sub-network is equipped with an adapter—1\(\times\)1 convolutional layer—tailored to its current context. This 1\(\times\)1 convolutional adapter transforms the combined features into the final output, thereby enriching the feature used for classification with the accumulated knowledge from the predecessors.

\textbf{A gap-aware layer scaling mechanism} adjusts the complexity of the SSNet feature extractor's residual blocks. Following~\cite{cl3}, we scale the blocks based on the minimum angle error tolerance.  For the smallest error tolerance of 5 degrees, we reduce the residual blocks from five to one because leveraging lateral connections allows sub-networks to share capacity. This manual scaling constrains the growth of the model parameters for more efficient SSL applications.

In this architecture, for each new learning task labeled as \(\tau_t\), we compute the real and imaginary parts of the complex STFT of audio segments to serve as input data. Concurrently, the CL-SSL approach preserves the integrity of the previously trained \(t-1\) sub-networks by freezing their parameters, preventing interference with newly acquired knowledge. The creator then creates a fresh  \(t^{{\text {th}}}\) sub-network dedicated to the current learning task, leveraging both the STFT features and the task metadata—such as task id—to discern previously unidentified acoustic configurations. What sets the \(t^{{\text {th}}}\) sub-network apart is its ability to access the shared memory of its predecessors through lateral connections, integrating all previous knowledge. 

When it comes to inference application, the CL-SSL model selects the appropriate  \(t^{{\text {th}}}\) sub-network based on the specific task at hand for accurate and reliable evaluation. This approach ensures that our model remains adaptable and quick to respond in diverse acoustic settings, showcasing the potential of continual learning strategy in the real-world environments of SSL.

\section{Experiments}
\label{sec:exp}

\begin{table}[t]
\centering
\caption{Parameter settings of simulated audio}

\label{tab:rir}
\resizebox{\columnwidth}{!}{%
\begin{tabular}{cccc}
\hline
\textbf{Room Size  [L, W, H] } & \textbf{Absorption} & \textbf{Tmax}    & \textbf{Num\_images} \\ \hline 
[10,8,5], [5,4,3], [4,2,2]         & 0.2-0.8                 & 0.2-0.6 & [10,10,10]       \\ \hline
\end{tabular}%
}

\end{table}

\subsection{Dataset}
We evaluated the proposed method on both simulated and real-world datasets, using two microphones to localize the direction of arrival within a 180-degree azimuth range.

\textbf{Simulated dataset:} We create our simulated data based on the LibriSpeech~\cite{librispeech}, which comprises single-channel speech recordings sampled at 16 kHz. We have curated a selection of 100 speakers. This selection balances gender representation with 47 male and 53 female speakers and encompasses a total of 1,511 audio files (4.96 hours). We enhance the original single-channel speech data by creating two-channel versions with gpuRIR~\cite{gpurir}, arranging two linear microphones at distances ranging from 5 to 9 cm. The gpuRIR is a library for room impulse response (RIR) simulation using the image source method. The details of the simulated room environments are specified in Table \ref{tab:rir}. We first simulate the microphone center anywhere inside the room. Then we randomly choose a distance such that the source is inside the room to finally simulate the RIR. Based on the RIR, we generate the mixture data for each original single-channel speech data every 5 degrees from 0 to 180 degrees, which expands the audio files 37 times. 

\textbf{Real-world dataset:} We also tested our proposed method on the LOCATA dataset~\cite{locata} following the setting in~\cite{fu1}. The room size was \(7.1\times9.8\times3\) m, with a reverberation time of 0.55 s. We utilized microphones 6 and 9 from the DICIT array~\cite{woz}, spaced 8 cm apart. White, babble, and factory noises from the NOISEX-92~\cite{NOISEX92} dataset were employed as noise sources. The models, trained on a noise-augmented simulated dataset, were directly evaluated on the LOCATA dataset using the same microphone configuration. Various signal-to-noise ratio (SNR) levels were applied to showcase the CL-SSL performance on real-world datasets.
\subsection{Metrics}
We utilize two primary metrics for evaluating our models: mean absolute error (MAE) and accuracy (ACC). They are all averaged on all learned tasks. MAE measures the average magnitude of the errors in a set of predictions, without considering their direction. On the other hand, ACC is used to determine the percentage of predictions that are exactly correct or within a certain degree of tolerance. We assess accuracy at three tolerances: Within a 5-degree, 10-degree, and 15-degree error margin. These common tiers of tolerance allow us to understand the model's precision and practical effectiveness.
\begin{figure}[!t]
   \centering
  \begin{minipage}[tbp]{\linewidth}
  \centering
  \includegraphics[width=\linewidth]{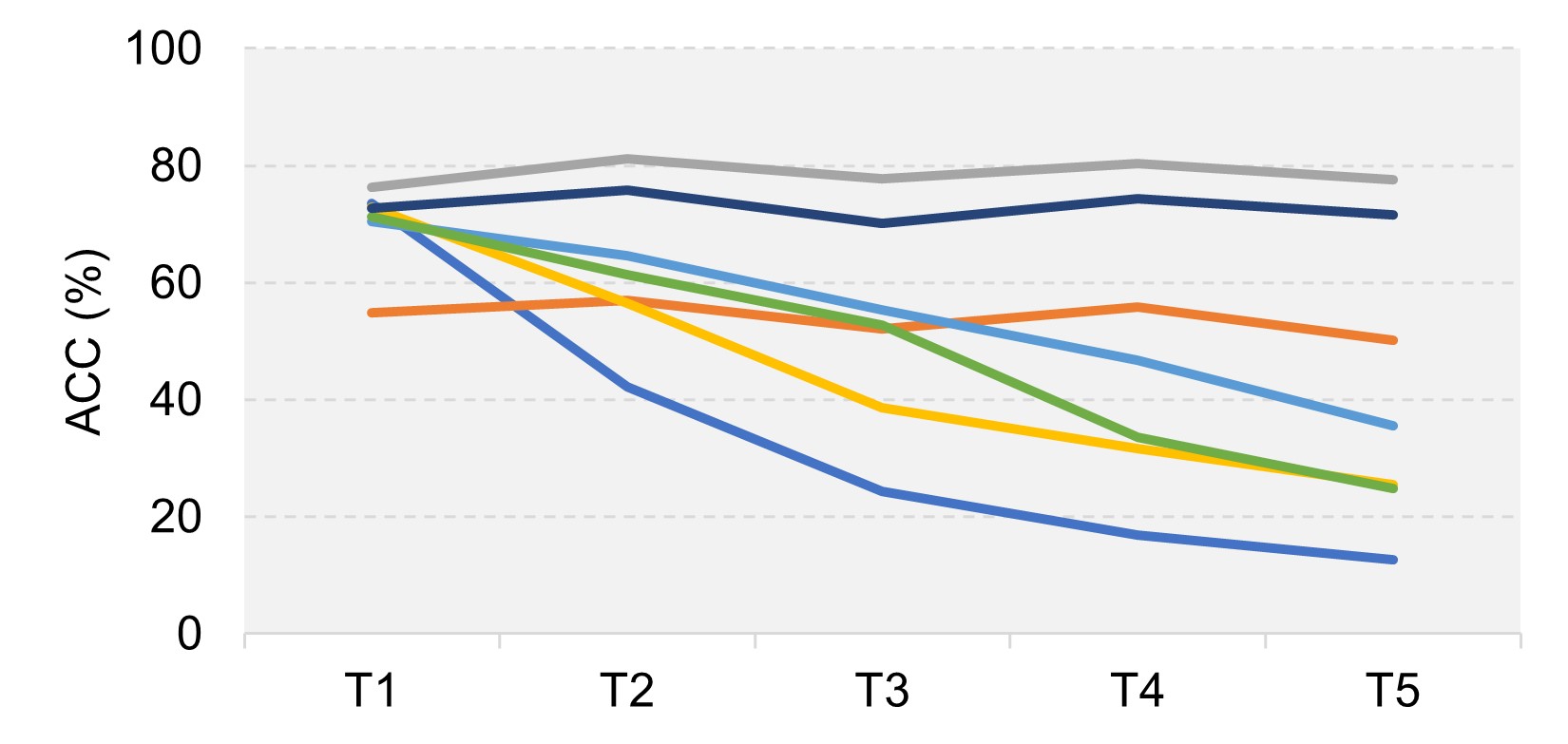}
  \centerline{(a) ACC \(\pm 5\)}
  \end{minipage}
\hfill
  \begin{minipage}[tbp]{\linewidth}
  \centering
    \includegraphics[width=\linewidth]{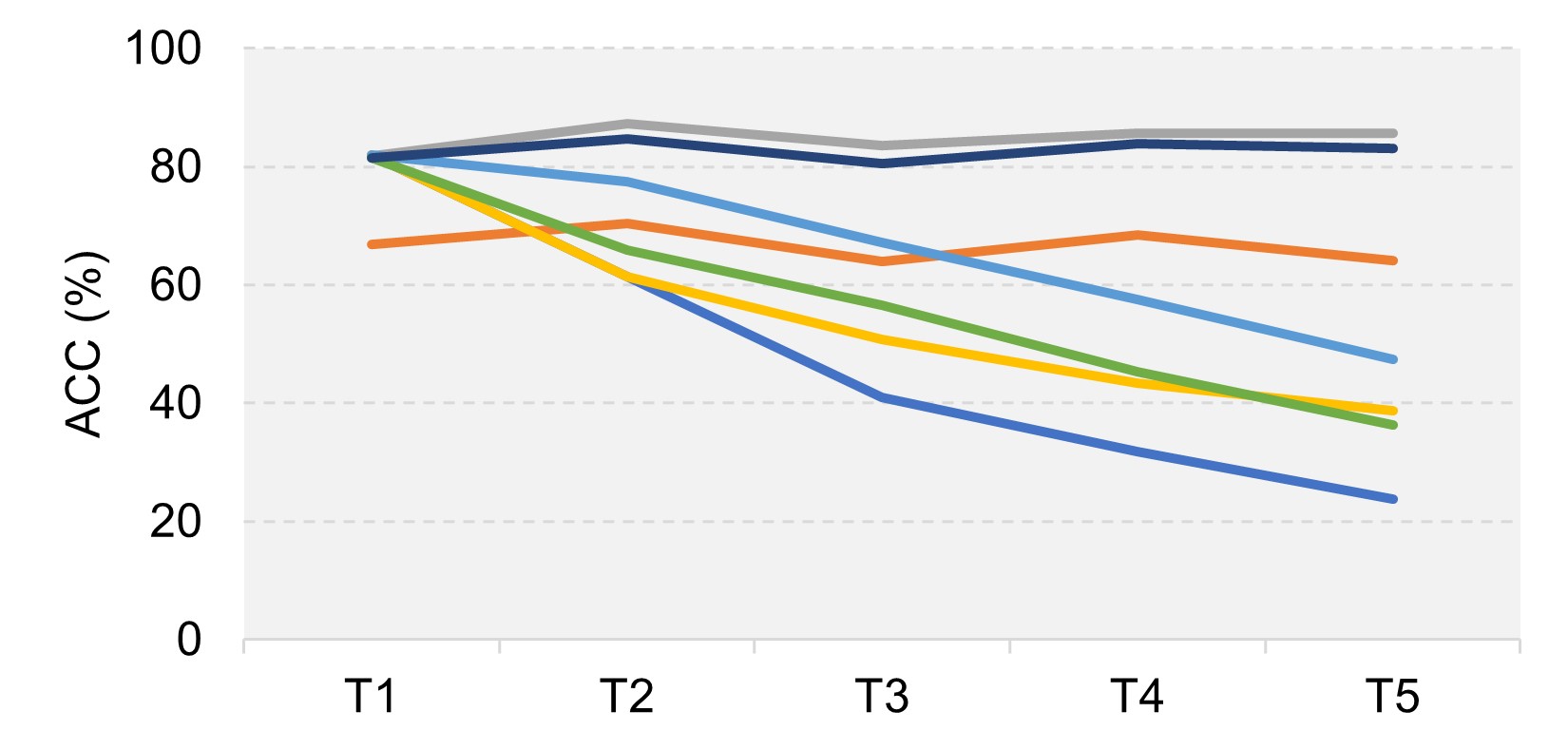}
  \centerline{(b) ACC \(\pm 10\)}    
  \end{minipage}
\hfill
  \begin{minipage}[tbp]{\linewidth}
  \centering
    \includegraphics[width=\linewidth]{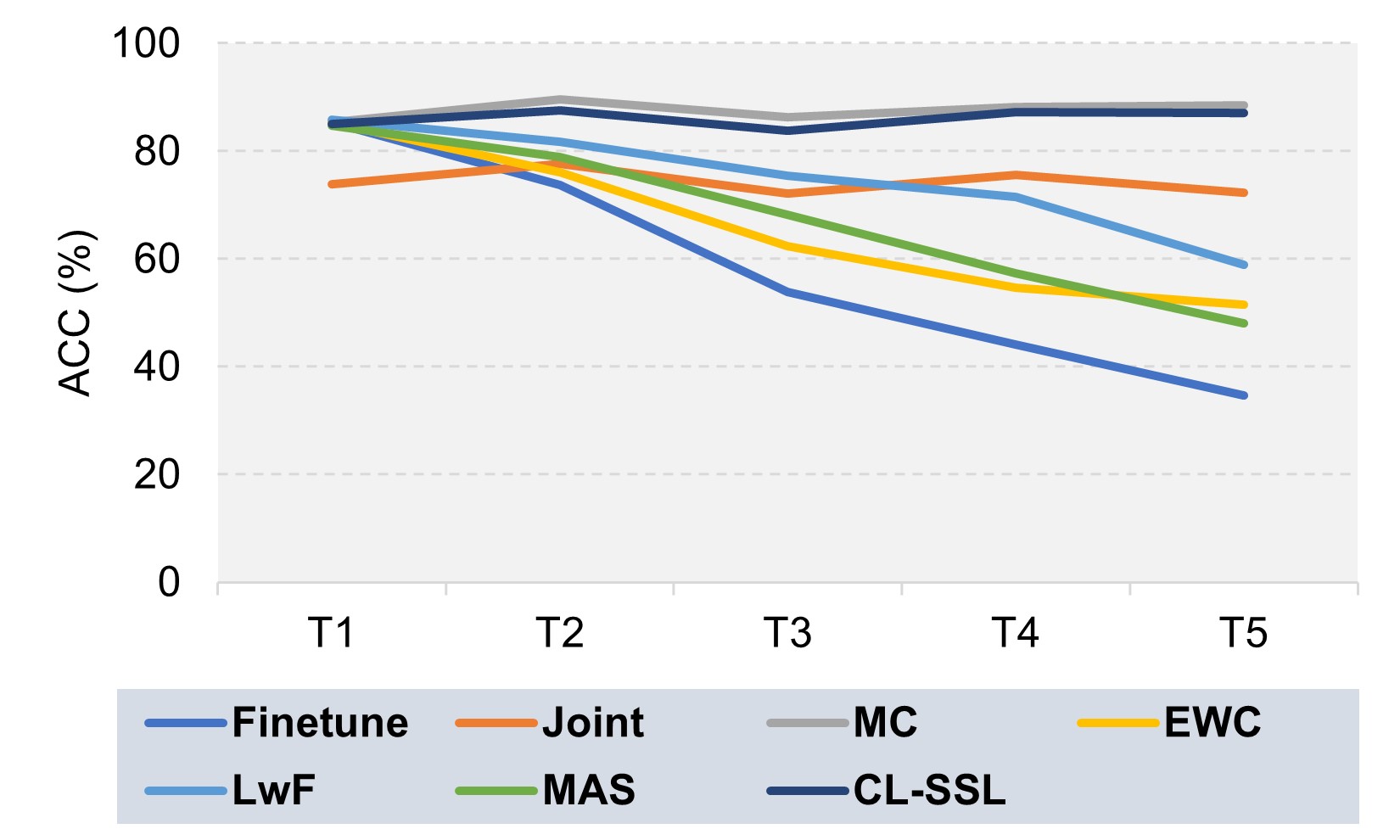}
  \centerline{(c) ACC \(\pm 15\)}    
  \end{minipage}
  \caption{Comparative performance in ACC (\%) of various SSL methods after learning each microphone spacing for three tolerances levels. The tasks T1-T5 are microphones distance from 5 to 9 cm.}
  \label{fig:acc}
\end{figure}

\subsection{Incremental learning setting}
In our incremental learning approach, the simulated dataset is divided into five parts based on microphone distances ranging from 5 to 9 cm. This setup frames the problem as a series of incremental learning tasks, each corresponding to a specific microphone spacing. For real-world data, noise levels are added using SNR values randomly selected from [-5dB, 0dB, 5dB, 10dB, Clean], representing different acoustic environments. The task order is randomly shuffled to better reflect practical scenarios. We compare our approach with several baseline methods:
\begin{itemize}
    \item \textbf{Finetune:} Starting with a model pre-trained on the initial task, we sequentially fine-tune it on each subsequent task. This approach adapts the model incrementally but may be susceptible to forgetting previously learned information when moving to new tasks each time.
    \item \textbf{Joint:} A single model is trained on the entire dataset, encompassing all conditions. This method aims to learn a general representation that performs well across all tasks. However, it can not handle large domain mismatches.
    \item \textbf{Multi-Condition (MC):} This involves training separate models for each condition on a different subset of the data, with each model specializing in a particular task. {\it In incremental task setting, the performance can never surpass the MC as it already knows the whole target classes the first time and ensemble multiple models.}
    \item \textbf{Elastic Weight Consolidation (EWC)~\cite{ewc}:} is a continual learning method that uses a quadratic penalty to restrict changes to parameters critical for previous tasks. These critical parameters are identified using an approximation based on the Fisher Information Matrix.
    \item \textbf{Learning without Forgetting (LwF)~\cite{lwf}:} is a method that ensures previously learned knowledge is retained while training on new tasks. It uses knowledge distillation, where the model's predictions on old tasks are treated as ``teacher" during the new task training.
    \item \textbf{Memory Aware Synapses (MAS)~\cite{mas}:} is similar to EWC but identifies critical parameters by measuring their impact on output predictions. 
\end{itemize}

\subsection{Implementation details}
In our CL-SSL framework, each task is trained for up to 100 epochs, with early stopping triggered after 20 epochs without performance improvement. We use the AdamW optimizer with weight decay, a learning rate of 0.001, and the StepLR scheduler for adjustments during training. For STFT computations, we use a frame size of 32 ms, a hop size of 10 ms, and a frequency range of 100 Hz to 8000 Hz, with n\_fft set to 512. The Gaussian constant \( \sigma \) is set to 8 in our experiments.


\section{Results and Discussions}
\label{sec:res}
\subsection{Results on simulated dataset}
We first test the methods for SSL across varying microphone spacing and report the results in Fig. \ref{fig:acc} and Table \ref{tab:my-table}. For the non-CL methods, we observe the finetune method incrementally trained only performs well for the first task indicating a forgetting across learned tasks. It presents the highest MAE, suggesting that it might not preserve the nuances of SSL in different microphone spacing as effectively as the other methods in Table \ref{tab:my-table}. Observing the performance of joint training, we find that it is more resilient compared to finetune, yet it does not maintain a high performance similar to the MC method. It is possible that the single-model approach is unable to capture the nuances needed for localization at variable spacing. The MC method shows consistent performance across different spacings due to the training of separate models for each task. However, it demands five times more memory and training time, making it less practical for real-world applications. As the number of tasks grows, the memory and time requirements increase linearly.

For CL methods, all methods outperform finetuning, demonstrating their effectiveness in reducing forgetting. However, their performance remains inferior to that of joint training and MC. In addition, from Fig. \ref{fig:acc}, it is observed that the performance of EWC and MAS drops dramatically when the number of learning tasks increases. This gap may result due to inappropriate regularization, which limits the model's ability to learn new tasks while preserving prior knowledge. Among the reference CL methods, LwF performs better, likely because its use of the model's predictions of old tasks enables knowledge sharing across tasks.

Our proposed CL-SSL shows a promising ability to obtain higher accuracy levels, particularly within the most challenging \(\pm 5\)-degree margin as shown in Fig. \ref{fig:acc}. It achieves comparable accuracy, nearly matching the best-performing method in the \(\pm 15\)-degree range in Table \ref{tab:my-table}. This indicates that our continual learning strategy effectively transfers knowledge from previous configurations to new spacing scenarios. By employing the layer scaling mechanism, we maintain a parameter count of just 1.6M—lower than LwF—while achieving significantly better performance. This highlights the proposed CL-SSL as a balanced solution between accuracy and model complexity, making it an efficient choice for SSL applications with varying microphone spacings.



\begin{table}[t]
\centering
\caption{Performance comparison in ACC (\%)  across different methods on the simulated dataset, highlighting MAE. The performance is calculated after all tasks are learned.}
\label{tab:my-table}
\resizebox{\columnwidth}{!}{%
\begin{tabular}{lccccc}
\hline
\textbf{Methods}         & \textbf{MAE}  & \textbf{ACC(\(\pm 5^\circ\))} & \textbf{ACC(\(\pm 10^\circ\))} & \textbf{ACC(\(\pm 15^\circ\))}  & \textbf{Params} \\ \hline 
Finetune         & 26.5 & 12.7  & 23.8    & 34.7  & 0.9M   \\ 
Joint           & 13.0 & 50.2  & 64.1  & 72.2   & 0.9M   \\ 
MC  & 7.6 & 77.6  & 85.6  & 88.4    & 4.5M   \\ \hline
EWC~\cite{ewc} & 23.5 & 25.4  & 38.7  & 51.4    & 0.9M \\
LwF~\cite{lwf} & 22.1 & 35.5  & 47.4  & 58.9    & 1.8M \\
MAS~\cite{mas} & 23.3 & 24.8  & 36.3  & 48.0    & 0.9M \\ 
\cellcolor[HTML]{C4D5EB}Proposed CL-SSL & \cellcolor[HTML]{C4D5EB}10.1 & \cellcolor[HTML]{C4D5EB}71.5  & \cellcolor[HTML]{C4D5EB}83.0  & \cellcolor[HTML]{C4D5EB}87.0   & \cellcolor[HTML]{C4D5EB}1.6M   \\ \hline
\end{tabular}%
}
\vspace{-3mm}
\end{table}

\subsection{Results on real-world dataset}

As discussed, we further evaluate the proposed CL-SSL method on a noisy real-world dataset. Noise is added based on SNR values randomly selected from [-5 dB, 0 dB, 5 dB, 10 dB, Clean], representing different acoustic environments, with task order randomly shuffled for practicality. The results compared with other baselines are summarized in Table~\ref{tab:noise-table}. We observe better performance across all methods on the LOCATA dataset due to its more stable environment, including consistent room size, microphone center, and room impulse response. The finetune method remains the lower bound among non-CL methods, indicating that catastrophic forgetting negatively impacts performance across varied acoustic environments due to domain mismatches. Joint training achieves relatively better results on real-world data than on the simulated data (Table~\ref{tab:my-table}), likely because optimizing for diverse acoustic environments is easier than for varying microphone configurations. The MC method, serving as the upper bound, performs well but at the cost of significantly higher memory usage.

Among the CL baselines, EWC, LwF, and MAS show comparable performance and outperform the finetune method, suggesting that these approaches effectively mitigate forgetting while balancing adaptation to new tasks. Our proposed CL-SSL method demonstrates less forgetting compared to other CL methods and performs closely to the MC method. When compared to joint training, often regarded as the upper bound in many applications, CL-SSL achieves comparable results despite not having access to all tasks in advance. This suggests that our method has greater potential to perform well in unseen acoustic environments. Notably, in the most challenging \(\pm 5\)-degree scenario, CL-SSL even surpasses joint training. This advantage may stem from its ability to adaptively scale parameters and focus on critical features, enabling better fine-grained localization under strict error tolerances.

\begin{table}[t]
\centering
\caption{Performance comparison in ACC (\%) across different methods on the real-world dataset, highlighting MAE. The performance is calculated after all tasks are learned.}
\label{tab:noise-table}
\resizebox{\columnwidth}{!}{%
\begin{tabular}{lccccc}
\hline
\textbf{Methods}         & \textbf{MAE}  & \textbf{ACC(\(\pm 5^\circ\))} & \textbf{ACC(\(\pm 10^\circ\))} & \textbf{ACC(\(\pm 15^\circ\))}  & \textbf{Params} \\ \hline 
Finetune         & 17.2 & 43.0  & 58.6    & 67.7  & 0.9M   \\ 
Joint           & 5.2 & 68.4  & 78.9  & 86.8   & 0.9M   \\ 
MC  & 4.3 & 78.9  & 89.4  & 94.7    & 4.5M   \\ \hline
EWC~\cite{ewc} & 9.3 & 47.3  & 68.4  & 81.5    & 0.9M \\
LwF~\cite{lwf} & 8.9 & 50.0  & 65.8  & 81.8    & 1.8M \\
MAS~\cite{mas} & 10.8 & 48.8  & 67.1  & 77.4    & 0.9M \\ 
\cellcolor[HTML]{C4D5EB}Proposed CL-SSL & \cellcolor[HTML]{C4D5EB}5.1 & \cellcolor[HTML]{C4D5EB}72.9  & \cellcolor[HTML]{C4D5EB}78.9  & \cellcolor[HTML]{C4D5EB}86.5   & \cellcolor[HTML]{C4D5EB}1.6M   \\ \hline
\end{tabular}%
}
\vspace{-3mm}
\end{table}

\subsection{Future Directions}
We foresee several avenues for future work. While we currently focus on two-microphone linear arrays to estimate the azimuth angle, we plan to expand our research to encompass more complex acoustic configurations~\cite{fu3} and to address elevation angle estimation for a full spatial representation. 

\section{Conclusions}
\label{sec:con}
In this study, we presented CL-SSL, an approach that advances the direction of arrival estimation within continual learning. CL-SSL synergizes task-specific network instantiation with progressive learning principles, addressing the diversity in microphone configurations with a novel scaling mechanism and lateral connections for knowledge retention. Demonstrated across simulated acoustic settings, CL-SSL consistently achieved high accuracy, underscoring its capacity for real-world applicability in dynamic auditory scenes. The model’s efficiency and adaptability suggest its potential as a foundational framework for future research in sound localization and acoustic signal processing, encouraging the development of sophisticated, scalable audio analysis technologies.

\clearpage
\balance
\bibliographystyle{IEEEbib}
\bibliography{icme2025references}

\end{document}